\documentclass[review,times,authoryear]{elsarticle}
\usepackage{graphicx}
\usepackage{amssymb}
\usepackage{multirow}
\usepackage{lineno}
\begin{document}

\title{Thermoelasticity of Fe$^{3+}$- and Al-bearing bridgmanite}

\author[spa]{Gaurav Shukla}\corref{cor1}
\ead{shuk0053@umn.edu}
\author[EPFL]{Matteo Cococcioni}
\author[spa,cems]{Renata M. Wentzcovitch}

\cortext[cor1]{Corresponding author}
\address[spa] {School of Physics and Astronomy, University of Minnesota, Minneapolis, Minnesota, USA}
\address[EPFL]{Theory and Simulation of Materials (THEOS), \'{E}cole polytechnique f\'{e}d\'{e}rale de Lausanne, Station 12, CH-1015 Lausanne, Switzerland}
\address[cems]{Department of Chemical Engineering and Materials Science, University of Minnesota, Minneapolis, Minnesota, USA}

\date{\today}
\setlength{\oddsidemargin}{-10pt}

\begin{abstract}
We report \textit{ab initio} (LDA + U$_{sc}$) calculations of thermoelastic properties of ferric iron (Fe$^{3+}$)- and  aluminum (Al)-bearing bridgmanite (MgSiO$_3$ perovskite), the main Earth forming phase, at relevant pressure and temperature conditions and compositions. Three coupled substitutions, namely,  [Al]$_{Mg}$-[Al]$_{Si}$, [Fe$^{3+}$]$_{Mg}$-[Fe$^{3+}$]$_{Si}$, and [Fe$^{3+}$]$_{Mg}$-[Al]$_{Si}$ have been investigated. Aggregate elastic moduli and sound velocities are successfully compared with limited experimental data available. In the case of the [Fe$^{3+}$]$_{Mg}$-[Fe$^{3+}$]$_{Si}$ substitution, the high-spin (S=5/2) to low-spin (S=1/2) crossover in [Fe$^{3+}$]$_{Si}$ induces a volume collapse and elastic anomalies across the transition region. However, the associated anomalies should disappear in the presence of aluminum in the most favorable substitution, i.e., [Fe$^{3+}$]$_{Mg}$-[Al]$_{Si}$. Calculated elastic properties along a lower mantle model geotherm suggest that the elastic behavior of bridgmanite with simultaneous substitution of Fe$_{2}$O$_3$ and Al$_{2}$O$_3$ in equal proportions or with Al$_{2}$O$_3$ in excess should be similar to that of (Mg,Fe$^{2+}$)SiO$_3$ bridgmanite. Excess Fe$_{2}$O$_3$ should produce elastic anomalies though.

\end{abstract}


\maketitle
\section{Introduction}

Aluminum (Al) and iron (Fe)-bearing bridgmanite (br), (Mg,Fe,Al)(Si,Fe,Al)O$_3$ perovskite, is the dominant mineral of the Earth's lower mantle along with (Mg,Fe)O ferropericlase (fp), CaSiO$_3$ perovskite, and (Mg,Fe,Al)(Si,Fe,Al)O$_3$ post-perovskite (PPv). In order to unravel the composition and thermal structure of the lower mantle, a very good estimate of thermal and elastic properties of these constituent minerals is required.  In spite of considerable experimental \citep{Sinnelnikov,Fiquet,Andrault,Jackson04,Sinogeikin,Jackson05b,Li05,Murakami07,Lundin,Ballaran,Chantel12,Murakami12,Dorfman} and computational effort \citep{Karki99,Karki01,Kiefer,Wentzcovitch04,Wentzcovitch09,Wu09,Wu11,Tsuchiya13,Wu13,Zhang13,Wu14,Shukla15,Wang15,Zhang16}, the available data on high pressure and high temperature elastic properties is still quite limited due to lack of sufficient understanding of the physics and chemistry of these minerals under extreme pressure and temperature conditions typical of the Earth's lower mantle. It has been shown that the iron in (Mg,Fe)O ferropericlase undergoes a pressure induced crossover from a high-spin (S=2) to a low-spin (S=0) state \citep{Badro03,Goncharov,Tsuchiya06}, which noticeably affects elastic properties \citep{Crowhurst08,Marquardt09,Wentzcovitch09,Wu09,Antonangeli11,Wu13,Wu14}. In the case of Al- and Fe-bearing bridgmanite, the elasticity data base is even more limited due to uncertainties associated with the possible coexistence of ferrous (Fe$^{2+}$) and ferric (Fe$^{3+}$) iron and their pressure induced state changes. Moreover, the uncertainties associated with the site occupancy of Al  and Fe in the perovskite structure make elasticity measurements and calculations of Al- and Fe-bearing bridgmanite even more challenging. By now it is believed that Fe$^{2+}$ occupies the A-site ([Fe$^{2+}$]$_{Mg}$), while Fe$^{3+}$ can occupy A- ([Fe$^{3+}$]$_{Mg}$) and/or B-site ([Fe$^{3+}$]$_{Si}$) of the perovskite structure  \citep{Badro04,Li04,Jackson05a,Li06,Stackhouse07,Lin08,Bengtson09,Dubrovinsky10,Hsu10,Hsu11,Fujino12,Hummer12,Lin12,Lin13,Tsuchiya13,Caracas14,Sinmyo14,Mao15}. In the entire lower mantle pressure-range, [Fe$^{2+}$]$_{Mg}$ remains in the HS state (S=2) but undergoes a pressure induced lateral displacement \citep{Bengtson09,Hsu10}, which results in a state with increased iron M\"{o}ssbauer quadrupole splitting (QS) \citep{McCammon,Bengtson09,Hsu10,Lin12,Lin13,McCammon13,potapkin13,kupenko14,Shukla15}. On the other hand, [Fe$^{3+}$]$_{Si}$ goes from the HS (S=5/2) to the LS (S=1/2) state and [Fe$^{3+}$]$_{Mg}$ remains in the HS (S=5/2) state throughout the lower mantle pressure region \citep{Catalli10,Catalli11,Fujino12,Fujino14,Hsu11,Lin12,Lin13,Tsuchiya13,Mao15,Shukla15b,Xu15}. For Al-bearing bridgmanite, experimental observations suggest that Al enters the lattice through coupled substitutions as [Al]$_{Mg}$-[Al]$_{Si}$ \citep{Andrault,Jackson04,Jackson05b}. In the case of simultaneous substitution of Al and Fe$^{3+}$ in bridgmanite, \citet{Hsu12} reported that [Fe$^{3+}$]$_{Mg}$-[Al]$_{Si}$ substitution is energetically favored with [Fe$^{3+}$]$_{Mg}$ remaining in the HS state (S=5/2) in the entire lower mantle pressure range.  

Based on X-ray emission spectroscopy (XES) and Synchrotron M\"{o}ssbauer spectroscopy (SMS), it has been suggested  that [Fe$^{3+}$]$_{Si}$ in Fe$^{3+}$-bearing bridgmanite undergoes a pressure induced HS to LS crossover in the pressure range 48-63 GPa \citep{Catalli10} and 18-25 GPa \citep{Lin12,Mao15} for the samples containing $\sim$10 mol.\% and $\sim$0.01 mol.\% Fe$_2$O$_3$, respectively.  Using Synchrotron X-ray diffraction data, \citet{Mao15} recently observed a noticeable volume collapse across the crossover region and this observation is consistent with the previous \textit{ab initio} predictions \citep{Hsu11,Tsuchiya13}.  Similarly to the spin crossover in (MgFe)O ferropericlase \citep{Jackson06,Crowhurst08,Marquardt09,Antonangeli11,Wentzcovitch09,Wu09,Wu13}, this volume collapse should cause elastic anomalies across the spin crossover region. These anomalies may have significant impact on the lower mantle properties \citep{Wu14}. Owing to extremely high pressure and temperature conditions, experimental observations of elastic anomalies of Fe$^{3+}$-bearing bridgmanite are still not available. Using \textit{ab initio} molecular dynamics (MD) simulations, \citet{Zhang16} recently calculated thermoelastic properties of [Fe$^{3+}$]$_{Mg}$-[Al]$_{Si}$-bearing bridgmanite. Their calculations were limited to few pressures and temperatures, did not include effect of strong correlation on Fe$^{3+}$, and conclusions about lower mantle composition were fitting parameters dependent. They also did not consider the effect of iron spin crossover on elastic properties. Here we present an \textit{ab initio} quasiharmonic calculation of thermoelastic properties of Al- and Fe$^{3+}$-bearing bridgmanite. This should help to clarify the potential consequences of the spin crossover of [Fe$^{3+}$]$_{Si}$ for lower mantle properties. 

\section{Computational details}

\begin{figure*}\centering
\includegraphics[width=15cm]{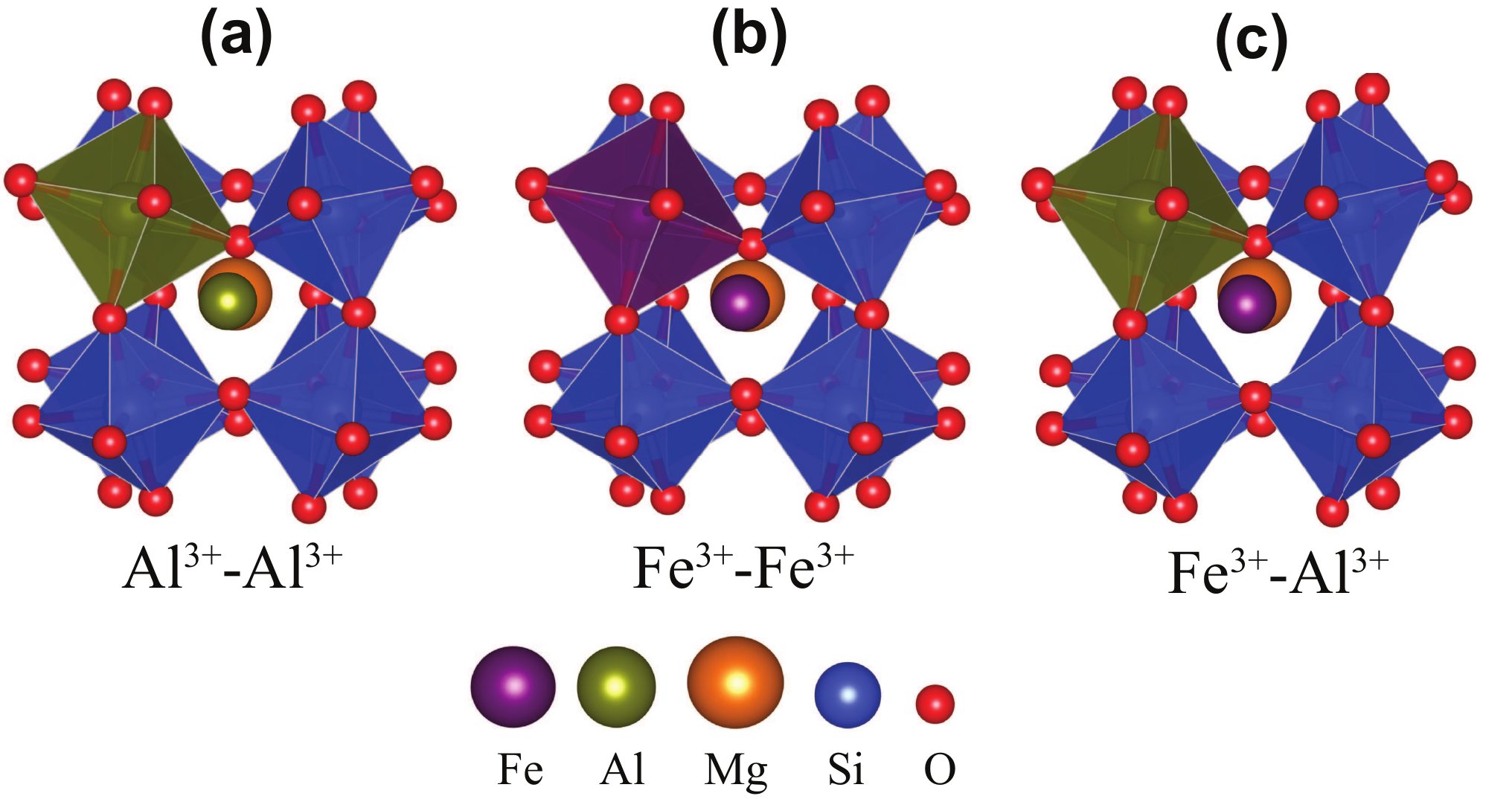}
\caption{(a) Atomic structure  of  Al- and Fe$^{3+}$-bearing bridgmanite with coupled substitutions (a) [Al]$_{Mg}$-[Al]$_{Si}$, (b) [Fe$^{3+}$]$_{Mg}$-[Fe$^{3+}$]$_{Si}$, and (c) [Fe$^{3+}$]$_{Mg}$-[Al]$_{Si}$.  Fe,Al, Mg, Si, and O are represented as purple, green, orange, blue, and red sphere, respectively. }
\label{fig_ferric1}
\end{figure*}

Here we have used density functional theory augmented by the self- and structurally consistent Hubbard type correction (DFT + U$_{sc}$ method)\citep{Cococcioni,Kulik} within the local density approximation (LDA) \citep{Ceperley-Alder80} to address the strong Coulomb correlation effects among iron $3d$ electrons. Three coupled substitutions of Al and Fe$^{3+}$ (shown in Figure~\ref{fig_ferric1}): (Mg$_{1-x}$Al$_x$)(Si$_{1-x}$Al$_x$)O$_3$ (Al-br), (Mg$_{1-x}$Fe$_x$)(Si$_{1-x}$Fe$_x$)O$_3$ (Fe$^{3+}$-br), and (Mg$_{1-x}$Fe$_x$)(Si$_{1-x}$Al$_x$)O$_3$ bridgmanite (Fe$^{3+}$-Al-br) for $x$ = 0.125 have been investigated. U$_{sc}$ values used here have been reported previously by \citet{Hsu11}. Elastic moduli and density for $0<x<0.125$ have been linearly interpolated using calculated results for $x$ = 0 and $x$ = 0.125.  All calculations have been performed in a 40-atom super-cell ($x = 0.125$). Ultrasoft pseudo-potentials \citep{Vanderbilt90} have been used for Al, Fe, Si, and O. For Mg, a norm-conserving pseudo-potential generated by von Barth-Car's method has been used. A detailed description of these pseudo-potentials has been reported by \citet{Umemoto08}. The plane-wave kinetic energy and charge density cut-off are 40 Ry and 160 Ry, respectively. For all three configurations mentioned above, the electronic states have been sampled on a shifted $2\times2\times2$ k-point grid \citep{Monkhorst-Pack76}. The equilibrium geometry at arbitrary pressure has been obtained using the variable cell-shape damped molecular dynamics (VCS-MD) \citep{Wentzcovitch91,Wentzcovitch93}. Vibrational effects have been addressed within the quasiharmonic approximation (QHA) \citep{Carrier07,Wallace72}. Vibrational density of states (VDOS) have been computed using the density functional perturbation theory (DFPT) \citep{Baroni01} within the LDA + U$_{sc}$ implementation  \citep{Floris11} for the exchange-correlation energy functional. 

Aggregate elastic moduli and acoustic velocities were calculated for [Al]$_{Mg}$-[Al]$_{Si}$, Fe$^{3+}$]$_{Mg}$-[Fe$^{3+}$]$_{Si}$, and  [Fe$^{3+}$]$_{Mg}$-[Al]$_{Si}$ substitutions using a semi-analytical method (SAM) \citep{Wu11}. For this purpose, equilibrium structures were obtained at 10-12 pressure points in the relevant pressure-range and dynamical matrices at each pressure point were calculated using DFPT in a $2\times2\times2$ q-point grid. Thus, obtained force constants from dynamical matrices were interpolated in a $8\times8\times8$ q-point grid to obtain VDOS. The static elastic coefficients, $C^{static}_{ijkl}$, were obtained at each pressure by applying small $\pm$1\% strains to the equilibrium structure, relaxing internal degrees of freedom, computing internal stresses and extracting them from the stress-strain relation $\sigma_{ij} = \sum C_{ijkl}\epsilon_{kl}$. This strain is small enough to maintain linearity in the stress-strain relation and large enough for good numerical accuracy in the calculation. Owing to the orthorhombic crystal symmetry, bridgmanite has nine independent elastic constants (i.e., C$_{11}$, C$_{22}$, C$_{33}$, C$_{44}$, C$_{55}$, C$_{66}$, C$_{12}$, C$_{13}$, C$_{23}$ in Voigt notation). Bulk (K) and shear modulus (G) have been estimated by computing Voigt-Reuss-Hill averages \citep{Watt76}. Using K and G along with the high temperature equation of states (for density, $\rho$), we obtained the pressure and temperature dependent acoustic velocities (compressional velocity, $V_P = \sqrt{\frac{K + \frac{4}{3}G}{\rho}}$, shear velocity, $V_S = \sqrt{\frac{G}{\rho}}$, and bulk velocity, $V_{\Phi} = \sqrt{\frac{K}{\rho}}$).   

\section{Results and discussion}

\subsection{Elasticity of [Al]$_{Mg}$-[Al]$_{Si}$-bearing bridgmanite}

\begin{figure*}\centering
\includegraphics[width=16cm]{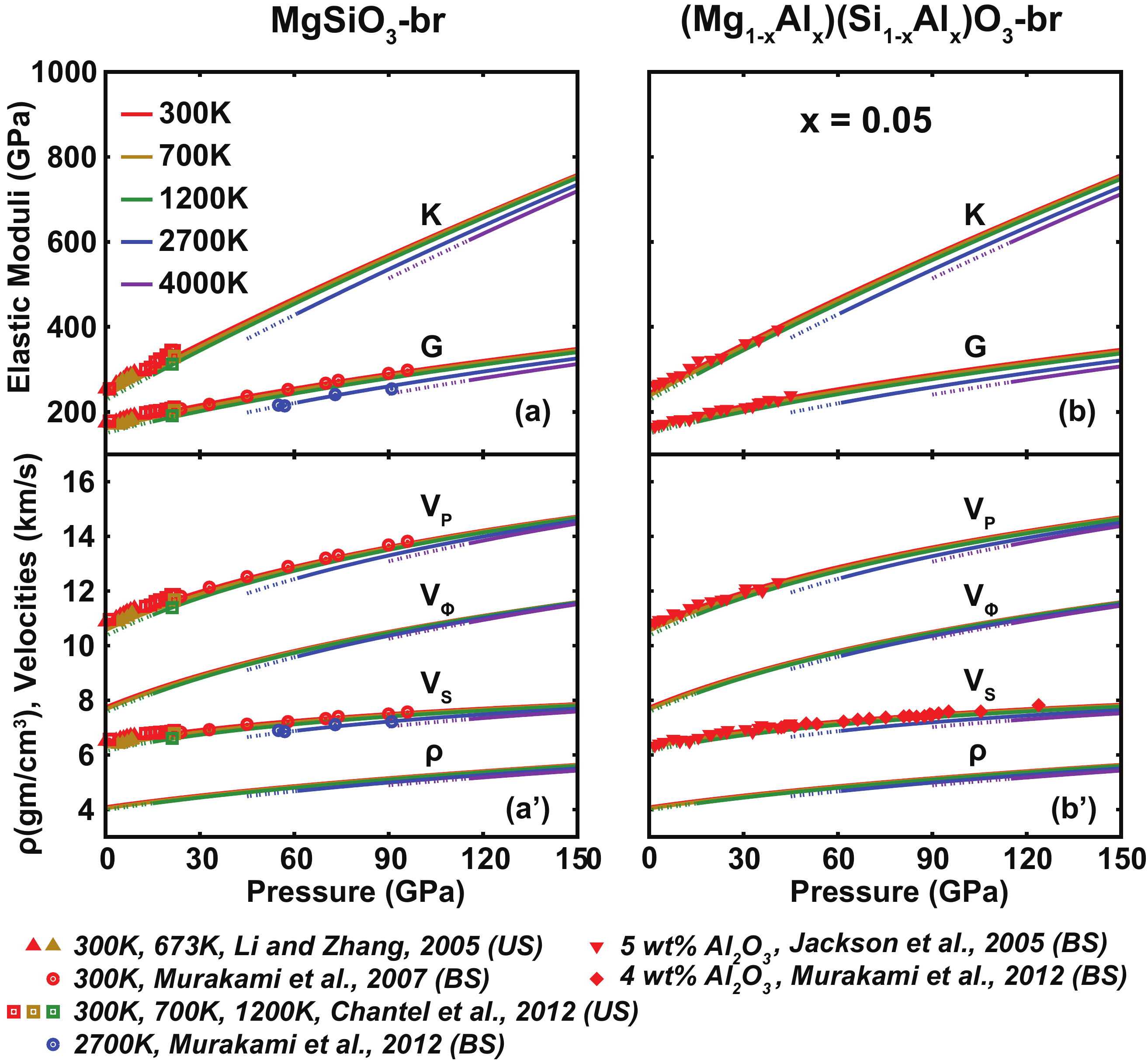}
\caption{Pressure and temperature dependence of elastic moduli, acoustic velocities, and density for (a,a') pure MgSiO$_3$ and (b,b')  (Mg$_{0.95}$Al$_{0.05}$)(Si$_{0.95}$Al$_{0.05}$)O$_3$ bridgmanite. Results for pure MgSiO$_3$ are taken from our previous study \citep{Shukla15}. Our calculated results (lines) are compared with available experimental data (symbols). Solid (dashed) lines represent \textit{ab initio} results within (outside) the validity of quasi-harmonic approximation. BS: Brillouin scattering, US: ultrasonic technique.}
\label{fig_ferric3}
\end{figure*}

Our calculated aggregate elastic moduli ($K$ and $G$), acoustic velocities ($V_P$, $V_S$, and $V_{\Phi}$), and density for pure MgSiO$_3$ and Al-bearing bridgmanite are shown in Figure~\ref{fig_ferric3}. Results of pure MgSiO$_3$ shown here have been taken from our previous study \citep{Shukla15}. Our calculated high temperature results, especially shear moduli and shear velocities at 2700 K, are in good agreement with  experimental data \citep{Li05,Chantel12,Murakami12} (Figure~\ref{fig_ferric3}a and \ref{fig_ferric3}a$'$), which shows that \textit{ab initio} methods used here provide a robust estimate for elastic properties of minerals at pressure and temperature conditions typical of the Earth's lower mantle. Our results for Al-br with $x = 0.05$ at 300 K also compare well with the room temperature data by \citet{Jackson05b} and \citet{Murakami12} for Al-bearing samples containing $\sim$5 wt\% and $\sim$4 wt\% of Al$_2$O$_3$, respectively (Figure~\ref{fig_ferric3}b and \ref{fig_ferric3}b$'$). For Al-br, calculated pressure derivatives of elastic moduli, $K'$ and $G'$, are 4.06 and 1.85, respectively. These values are fairly comparable with those reported by \citet{Jackson05b} (Table \ref{table:ambient-data}). In overall, our results of Al$_2$O$_3$ substitution at 300 K are consistent with previous experimental studies \citep{Jackson04,Jackson05b,Murakami12}.

\subsection{Elasticity of [Fe$^{3+}$]$_{Mg}$-[Fe$^{3+}$]$_{Si}$ and  [Fe$^{3+}$]$_{Mg}$-[Al]$_{Si}$-bearing bridgmanite}
Having successfully calculated the elastic properties for pure and Al-bearing bridgmanite, we now investigate the elastic consequences of the pressure induced spin crossover of [Fe$^{3+}$]$_{Si}$ in (Mg$_{1-x}$Fe$_x$)(Si$_{1-x}$Fe$_x$)O$_3$ bridgmanite (Fe$^{3+}$-br). For this purpose, we calculate elastic properties of the system in the mixed spin state (MS) by extending the approach developed by \cite{Wu13}. The elastic compliances $S^{ij}$ of Fe$^{3+}$-br in the MS state are given by (see Supporting Information for details)
\begin{eqnarray}
 S^{ij}V = nS^{ij}_{LS}V_{LS} + (1-n)S^{ij}_{HS}V_{HS} - \frac{1}{9}(V_{LS} - V_{HS})\frac{\partial n}{\partial P} 
\end{eqnarray}
for $i,j = 1-3$ and
\begin{eqnarray}
 S^{ii}V = nS^{ii}_{LS}V_{LS} + (1-n)S^{ii}_{HS}V_{HS}
\end{eqnarray}
for $i = 4-6$. Here $S^{ij}_{HS/LS}$ and $V_{HS/LS}$ are the elastic compliances and volume of pure HS/LS state, respectively. $n$ is the low-spin (LS) fraction of [Fe$^{3+}$]$_{Si}$ and is given by
\begin{eqnarray}
  n(P,T)= \frac{1}{1+\frac{m_{HS}(2S_{HS}+1)}{m_{LS}(2S_{LS}+1)}\exp\left\{\frac{\Delta G_{LS\rightarrow HS}}{xk_BT}\right\}},
\end{eqnarray}
 where S$_{HS/LS}$ is spin moment (S$_{HS}$ = 5/2, S$_{LS}$ = 1/2), m$_{HS/LS}$ is the orbital degeneracy (m$_{HS}$ = 1, m$_{LS}$ = 3) of HS/LS state, $k_B$ is the Boltzmann constant, and $\Delta G_{LS\rightarrow HS} = G^{stat+vib}_{LS} - G^{stat+vib}_{HS}$. Elastic coefficients $C_{ij}$ are obtained using the relation: $\textbf{C}_{ij} = \left[\textbf{S} ^{-1} \right]^{ij}$

\begin{figure*}\centering
\includegraphics[width=12cm]{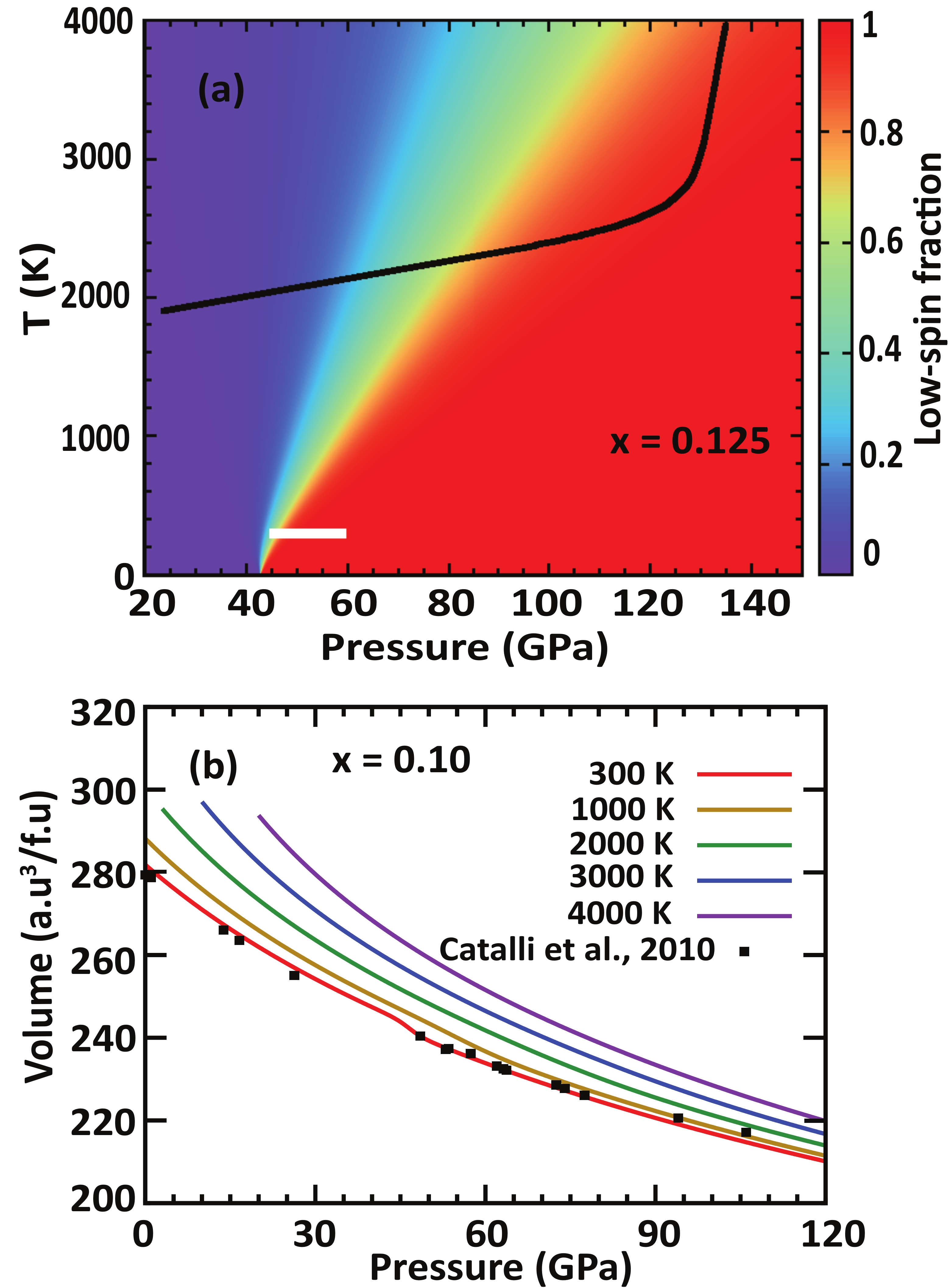}
\caption{ (a) Pressure and temperature dependence of low-spin fraction, n(P,T), of [Fe$^{3+}$]$_{Si}$ in (Mg$_{1-x}$Fe$^{3+}_{x}$)(Si$_{1-x}$Fe$^{3+}_{x}$)O$_3$ bridgmanite for $x = 0.125$. Solid black curve represents the lower mantle model geotherm by \citep{Boehler00}, while white bar represents the experimental pressure range at room temperature in which spin crossover of [Fe$^{3+}$]$_{Si}$ is complete \citep{Catalli10}. (b) Pressure dependence of calculated volume at several temperature for $x= 0.10$. Compression curve at 300 K (red) is compared with experimental data by \citet{Catalli10} (black squares).}
\label{fig_ferric4}
\end{figure*}

The pressure and temperature dependence of LS fraction of [Fe$^{3+}$]$_{Si}$, $n(P,T)$, in Fe$^{3+}$-br for $x = 0.125$ is shown in Figure~\ref{fig_ferric4}a. The spin crossover pressure range in our calculation is much broader than that reported by \citet{Tsuchiya13}. Our estimated transition pressure width at 300 K is about $\sim$8 GPa, in fairly good agreement with the experimental data \citep{Catalli10,Lin12,Mao15}, while the one reported by \citet{Tsuchiya13} is $<$2 GPa. This difference may be related to the use of different values of Hubbard U and different technique for VDOS calculations. The values of U used here were calculated self-consistently, while those reported by \citet{Tsuchiya13} were not. \citet{Tsuchiya13} used the finite displacement method \citep{Alfe09} to calculate the VDOS. This method disregards the calculation of dielectric constant tensor that leads to LO-TO splitting for polar materials. In this work, we have used DFPT + U method developed by \citet{Floris11} for computing dynamical matrices and phonon frequencies. This method has been successfully used to address vibrational properties and the [Fe$^{2+}$]$_{Mg}$ displacement transition in Fe$^{2+}$-bearing bridgmanite \citep{Shukla15}. Our calculated 300 K compression curve for Fe$^{3+}$-br with $x = 0.10$  also compare well with experimental data \citep{Catalli10} (Figure~\ref{fig_ferric4}b). Therefore, we believe that calculation of VDOS and its free-energy contribution using DFPT + U method is quite robust.

\begin{figure*}\centering
\includegraphics[width=16cm]{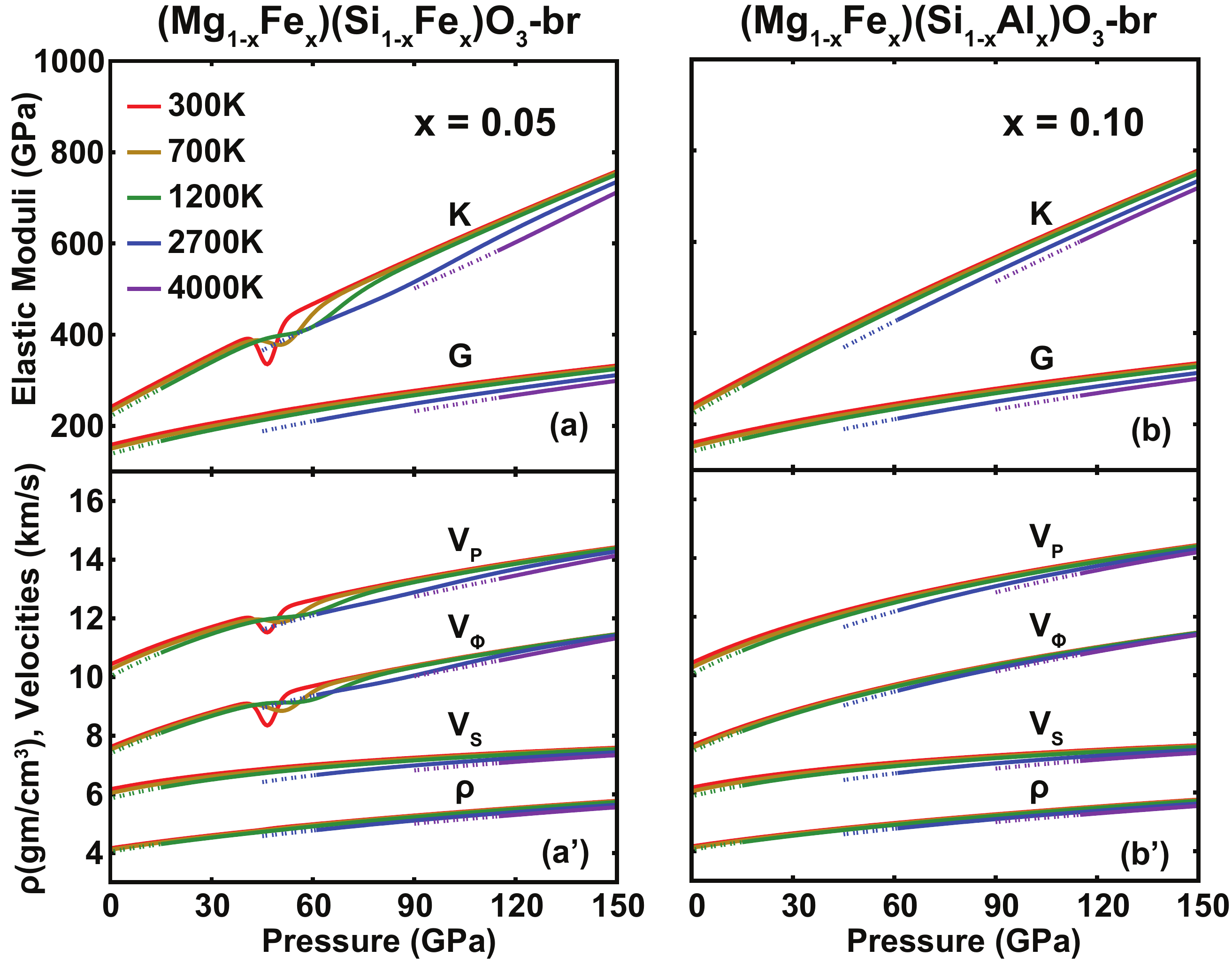}
\caption{Pressure and temperature dependence of elastic moduli (bulk modulus K, and shear modulus G) and acoustic velocities (compressional velocity V$_P$, shear velocity V$_S$, and bulk velocity V$_{\Phi}$) for (a,a') (Mg$_{0.95}$Fe$_{0.05}$)(Si$_{0.95}$Fe$_{0.05}$)O$_3$ and (b,b') (Mg$_{0.90}$Fe$_{0.10}$)(Si$_{0.90}$Al$_{0.10}$)O$_3$ bridgmanite. Solid (dashed) lines represent \textit{ab initio} results within (outside) the validity of quasi-harmonic approximation.}
\label{fig_ferric5}
\end{figure*}

Our results for elastic moduli, acoustic velocities, and density of Fe$^{3+}$-br with $x = 0.05$ are shown in Figure~\ref{fig_ferric5}a and ~\ref{fig_ferric5}a$'$. The spin crossover of [Fe$^{3+}$]$_{Si}$ produces anomalous softening in the bulk modulus in the crossover pressure region and this anomalous behavior is also reflected in compressional (V$_P$) and bulk velocities (V$_{\Phi}$). The strength of the anomaly depends on  temperature, volume difference ($\Delta V_{HS\rightarrow LS} = V_{LS} - V_{HS}$), and Gibb's free energy difference ($\Delta G_{HS\rightarrow LS} = G_{LS} - G_{HS}$). $\Delta V_{HS\rightarrow LS}$ for Fe$^{3+}$-br with $x$ = 0.01 is approximately $\sim$-0.15\% (Figure~\ref{fig_ferric4}b), which is in good agreement with previous \textit{ab initio} calculations \citep{Hsu11,Tsuchiya13} and is fairly comparable to experimental value $\sim$-0.2\% \citep{Mao15}. This volume reduction causes quite a significant softening in the bulk modulus ($\sim$12\% at 300 K), while the shear modulus increases only by a small amount ($\sim$0.5\% at 300 K) across the transition region. 

Effects of Al substitution on the spin crossover of Fe$^{3+}$ was previously studied by \citet{Hsu12} using \textit{ab initio} static calculations within the LDA + U$_{sc}$ approximation. The coupled substitution [Fe$^{3+}$]$_{Mg}$-[Al]$_{Si}$ was found to be energetically favorable. Therefore, we have chosen this substitution to investigate the elastic properties of bridgmanite due to simultaneous inclusion of Al and Fe. Elastic moduli and acoustic velocities for (Mg$_{1-x}$Fe$_{x}$)(Si$_{1-x}$Al$_{x}$)O$_3$ bridgmanite with $x$ = 0.10 are shown in Figure~\ref{fig_ferric5}b and Figure~\ref{fig_ferric5}b$'$. [Fe$^{3+}$]$_{Mg}$ is in the HS (S = 5/2) state in the entire lower mantle pressure range. Owing to the absence of any spin state crossover in [Fe$^{3+}$]$_{Mg}$-[Al]$_{Si}$ substitution, the elastic anomalies, otherwise present in the Fe$^{3+}$-bearing bridgmanite, disappear in this case.

\begin{table}
\caption{Calculated volume ($V$), elastic moduli ($K$, $G$), their pressure derivatives ($K', G'$), and acoustic velocities ($V_P$, $V_S$, $V_{\Phi}$) for Al- and Fe-bearing bridgmanite (br) at 0 GPa and 300 K. $^a$ See Table~1 in \citet{Shukla15} for detailed comparison of br and Fe$^{2+}$-br results with experimental data}
\centering
\label{table:ambient-data}
\resizebox{18cm}{!}{
\begin{tabular}[width=18cm]{ l l l l l l l l l l}
\hline
 & $V(\mathring{A}^3$) & $K$ (GPa) & $G$ (GPa) & $K'$ & $G'$ & $V_P$ & $V_S$ & $V_{\Phi}$ & References\\
\hline
\hline
br ($x$ = 0)$^a$ & 163.2 & 245.5 & 168.2 & 3.96 & 1.79  & 10.72 & 6.42 & 7.74 & \citet{Shukla15} \\
Fe$^{2+}$-br ($x$ = 0.125)$^a$ & 163.7 & 246.7 & 165.2 & 4.03 & 1.80 & 10.50 & 6.25 & 7.64 & \citet{Shukla15} \\
Al-br ($x$ = 0.125) & 164.1 & 242.5 & 162.7 & 4.06 & 1.85 & 10.62 & 6.32 & 7.72 & \textit{This study} \\
Al-br ($x \approx$ 0.0.5) & 163.26(1) & 252(5) & 165(2) & 3.7(0.3) & 1.7(0.2) & 10.75(5) & 6.35(5) & 7.85(5) &\citet{Jackson04,Jackson05b} \\
Fe$^{3+}$-br ($x$ = 0.125) & 168.0 & 233.6 & 143.1 & 4.05 & 1.88 & 9.98  & 5.80 & 7.40 & \textit{This study} \\
Fe$^{3+}$-Al-br ($x$ = 0.125) & 165.3 & 239.4 & 156.3 & 4.05 & 1.83 & 10.34 & 6.11 & 7.56 & \textit{This study}\\ 
\hline
\hline
\end{tabular}
}
\end{table}

\section{Geophysical significance}
Understanding the effect of Al and Fe substitution on the elastic properties of bridgmanite is essential to unravel the composition, thermal structure, nature of seismic heterogeneities, and dynamics of the Earth's lower mantle. Table~\ref{table:ambient-data} compares calculated volume, elastic moduli (bulk modulus, K, and shear modulus, G), their pressure derivatives (K$'$, G$'$), and acoustic velocities (V$_P$, V$_S$, and V$_{\Phi}$) at ambient pressure and temperature conditions for the most relevant substitutions.  Inclusion of 12.5 mol.\% of Al$_2$O$_3$ in bridgmanite changes volume, K, and G by $\sim$0.5\%, $\sim-$1.2\%, and $\sim-$3.5\%, respectively. This trend becomes more noticeable when the same amount (12.5 mol.\%) of Fe$_2$O$_3$ is substituted, which results in $\sim$3\%, $\sim-$5\%, and $\sim-$15\% change in volume, K, and G, respectively. It is important to note that the effect of Fe$_2$O$_3$ substitution is different from that of FeO. As seen in Table~\ref{table:ambient-data}, the presence of 12.5 mol.\% of FeO increases K by $\sim$0.5\%, while the same amount of Fe$_2$O$_3$ decreases it by $\sim$5\%. Simultaneous substitution of Fe$_2$O$_3$ and Al$_2$O$_3$ as [Fe$^{3+}$]$_{Mg}$-[Al]$_{Si}$ changes the volume, K, and G by $\sim$1.3\%, $\sim-$2.5\%, and $\sim-$7\%, respectively. These changes in volume and elastic moduli due to Al and Fe substitutions are also reflected in acoustic velocities. The incorporation of 12.5 mol.\% of Al$_2$O$_3$ in bridgmanite, changes V$_P$, V$_S$, and V$_{\Phi}$ by $\sim-$0.9\%, $\sim-$1.6\%, and  $\sim-$0.3\%, respectively, while the same amount of Fe$_2$O$_3$ changes these values by $\sim-$6.9\%, $\sim-$9.7\%, and $\sim-$4.4\%, respectively. For simultaneous substitution of Al$_2$O$_3$ and Fe$_2$O$_3$, the change in V$_P$, V$_S$, and V$_{\Phi}$ is $\sim-$3.5\%, $\sim-$4.8\%, and $\sim-$2.3\%, respectively.

Table~\ref{temp_deriv_K_G} compares  $\frac{dK}{dT}$ and $\frac{dG}{dT}$ for various Fe and Al substitutions at 300 K and 0, 60, and 120 GPa.
Results for pure, Fe$^{2+}$-, and [Fe$^{3+}$]$_{Mg}$-[Al]$_{Si}$-bearing bridgmanite are very similar. This observation suggests that the temperature dependence of elastic properties due to simultaneous inclusion of Al$_2$O$_3$ and Fe$_2$O$_3$ in bridgmanite would be nearly indistinguishable from those of FeO substitution with the same amount of iron. 

\begin{figure*}\centering
\includegraphics[width=18.0cm]{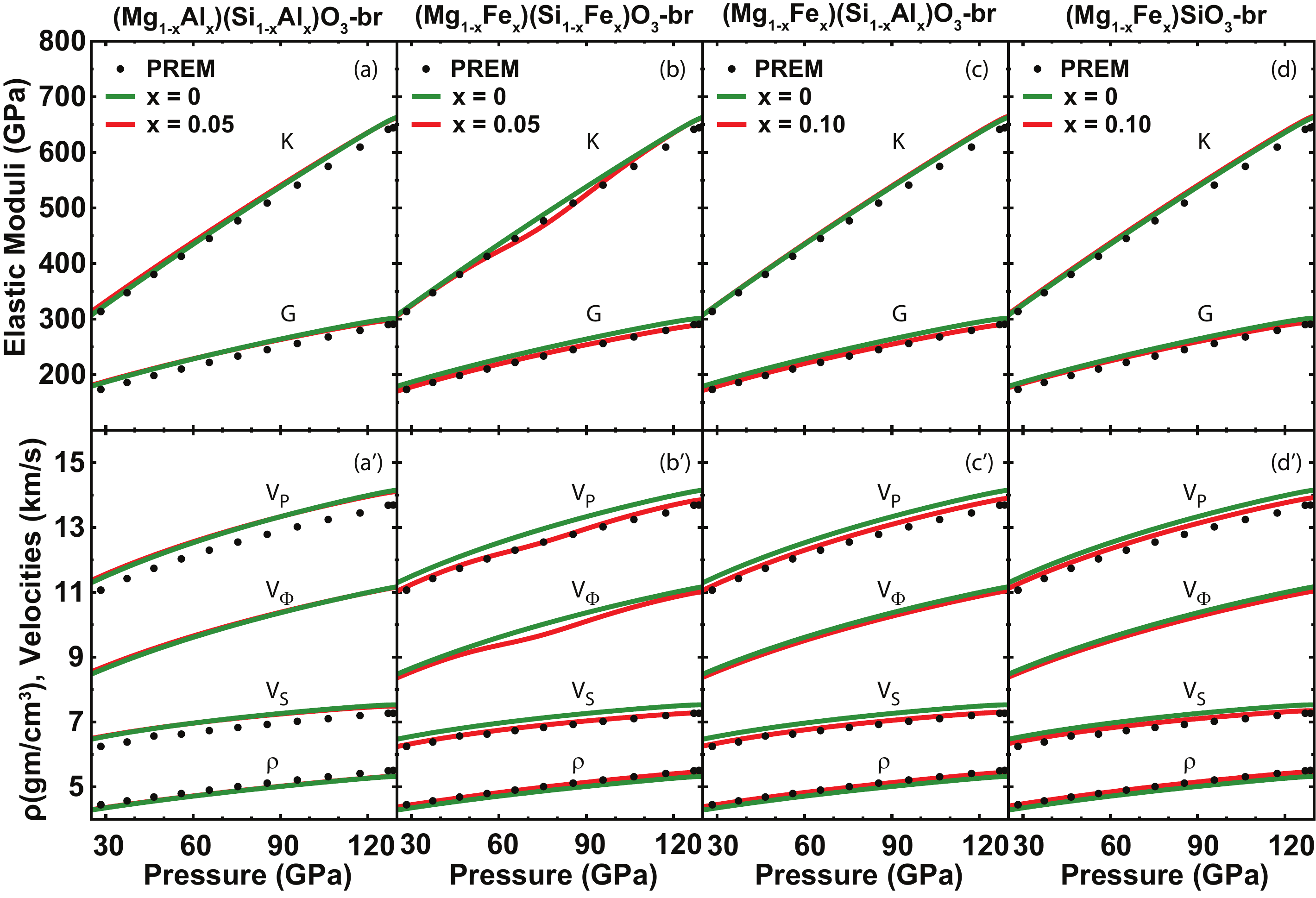}
\caption{Elastic moduli, acoustic velocities, and density along a model lower mantle  geotherm by \textit{Boehler,~2000} \citep{Boehler00} for (a,a') [Al]$_{Mg}$-[Al]$_{Si}$, (b,b') [Fe$^{3+}$]$_{Mg}$-[Fe$^{3+}$]$_{Si}$, (c,c') [Fe$^{3+}$]$_{Mg}$-[Al]$_{Si}$, and [Fe$^{2+}$]$_{Mg}$ substitution in MgSiO$_3$ bridgmanite. Calculated results (red lines) are compared with that of pure MgSiO$_3$ bridgmanite (green lines) and with the Preliminary Reference Earth Model (PREM) \citep{Dziewonski81} data (black points).}
\label{fig_ferric6}
\end{figure*}

Figure~\ref{fig_ferric6} shows calculated elastic moduli, acoustic velocities, and density along a model lower mantle geotherm \citep{Boehler00} for (Mg$_{0.95}$Al$^{3+}_{0.05}$)(Si$_{0.95}$Al$^{3+}_{0.05}$)O$_3$, (Mg$_{0.95}$Fe$^{3+}_{0.05}$)(Si$_{0.95}$Fe$^{3+}_{0.05}$)O$_3$, (Mg$_{0.90}$Fe$^{3+}_{0.10}$)(Si$_{0.90}$Al$^{3+}_{0.10}$)O$_3$, and (Mg$_{0.90}$Fe$^{2+}_{0.10}$)SiO$_3$ bridgmanite. These concentrations were chosen to represent the expected composition of bridgmanite in the lower mantle \citep{Irifune10}. Results are also compared with those of pure MgSiO$_3$ and with Preliminary Reference Earth Model (PREM) \citep{Dziewonski81}. The inclusion of Al$_2$O$_3$ does not seem to affect the elastic properties of bridgmanite significantly (Figure~\ref{fig_ferric6}a and \ref{fig_ferric6}a$'$). The anomalous softening due to spin crossover of [Fe$^{3+}$]$_{Si}$ is present in the [Fe$^{3+}$]$_{Mg}$-[Fe$^{3+}$]$_{Si}$ substitution (Figure~\ref{fig_ferric6}b and \ref{fig_ferric6}b$'$). 
However, these anomalies no longer exist with the simultaneous substitutions of Al and Fe as [Fe$^{3+}$]$_{Mg}$-[Al]$_{Si}$. This substitution mechanism, its elastic properties, and the likely low concentration of Fe$^{3+}$ in br \citep{Xu15}, might explain the lack of sizeable spin crossover induced elastic anomalies in the PREM model \citep{Dziewonski81}. In addition, the elastic properties of (Mg$_{0.90}$Fe$^{3+}_{0.10}$)(Si$_{0.90}$Al$^{3+}_{0.10}$)O$_3$ are very similar to those of (Mg$_{0.90}$Fe$^{2+}_{0.10}$)SiO$_3$ bridgmanite or to PREM velocities, except for the bulk modulus, K, and compressional velocity, V$_P$. Relatively larger K and V$_P$ compared to PREM values suggest that the lower mantle may accommodate a reasonable amount of ferropericlase, (MgFe)O, and CaSiO$_3$ perovskite, consistent with conclusions of previous first principles calculations \citep{Wentzcovitch04}.

\begin{table}
\caption{Calculated temperature derivatives of aggregate elastic moduli in unit of GPa/K at 300 K and 0, 60, and 120 GPa.}
\centering
\label{temp_deriv_K_G}
\resizebox{18cm}{!}{
\begin{tabular}[width=18cm]{ l c c c c c c}
\hline
 & $\frac{dK(0,300)}{dT}$ & $\frac{dK(60,300)}{dT}$  & $\frac{dK(120,300)}{dT}$  & $\frac{dG(0,300)}{dT}$  & $\frac{dG(60,300)}{dT}$  & $\frac{dG(120,300)}{dT}$ \\
\hline
\hline
br ($x$ = 0) & -0.0157 & -0.0093 & -0.0054 & -0.0178 & -0.0108  & -0.0077 \\
Fe$^{2+}$-br ($x$ = 0.125) & -0.0160 & -0.0097 & -0.0059 & -0.0178 & -0.0107 & -0.0077 \\
Al-br ($x$ = 0.125) & -0.0116 & -0.0095 & -0.0086 & -0.0144 & -0.0109 & -0.0095 \\
Fe$^{3+}$-br ($x$ = 0.125) & -0.0191 & -0.0088 & -0.0001 & -0.0205 & -0.0101 & -0.0066 \\
Fe$^{3+}$-Al-br ($x$ = 0.125) & -0.0165 & -0.0097 & -0.0054 & -0.0182 & -0.0106 & -0.0074 \\ 
\hline
\hline
\end{tabular}
}
\end{table}

\section{Conclusions}
We have presented \textit{ab initio} calculations of thermoelastic properties Al- and Fe$^{3+}$-bearing bridgmanite. Aggregate elastic moduli, acoustic velocities, and density  for three coupled substitutions, [Al]$_{Mg}$-[Al]$_{Si}$, [Fe$^{3+}$]$_{Mg}$-[Fe$^{3+}$]$_{Si}$, and [Fe$^{3+}$]$_{Mg}$-[Al]$_{Si}$ have been computed by combining the vibrational density of states and static LDA/LDA + U$_{sc}$ calculations within the quasiharmonic approximation. Results for pure MgSiO$_3$ and  Al-bearing bridgmanite compare very well with available experimental data \citep{Li05,Jackson05b,Chantel12,Murakami07,Murakami12}. Elastic anomalies due to spin crossover of [Fe$^{3+}$]$_{Si}$ in Fe$^{3+}$-bearing bridgmanite cease to exist in the presence of Al$_2$O$_3$. Elastic properties calculated along a model lower mantle  geotherm \citep{Boehler00} suggest that Al$_2$O$_3$ may be accommodated with Fe$_2$O$_3$ in a coupled substitution without altering the overall elastic properties of (Mg,Fe$^{2+}$)SiO$_3$ bridgmanite. 

\appendix
\section{Solid solution model for elasticity of mixed spin state (MS) of HS and LS state of [Fe$^{3+}$]$_{Si}$ in Fe$^{3+}$-bearing bridgmanite}
The molar Gibb's free energy for the system (under stress $\sigma$) in the MS state is given by
\begin{eqnarray}\label{G_MIX}
 G(P,T,n,\sigma) = nG_{LS}(P,T,\sigma)+(1-n)G_{HS}(P,T,\sigma) + G^{mix}(P,T,\sigma),
\end{eqnarray}
where $n(P,T,\sigma)$ is the low-spin (LS) fraction, T is temperature, P is pressure, $G_{LS/HS}$ is the Gibb's free energy for pure HS/LS state. $G^{mix}$ is the mixing free energy for ideal solid solution of HS and LS states of [Fe$^{3+}$]$_{Si}$ and is given by
\begin{eqnarray}
 G^{mix}(P,T) = xk_BT[nln(n) +(1-n)ln(1-n)],
\end{eqnarray}
where $x$ is Fe$_2$O$_3$ molar concentration, $k_B$ is Boltzmann constant.

The Gibb's free energy for pure HS/LS state is given as
\begin{eqnarray}
 G_{HS/LS}(T,V) &=&  G^{stat+vib}_{HS/LS}(T,V) + G^{mag}_{HS/LS}(T,V),
\end{eqnarray}
where $G^{stat+vib}_{HS/LS}(T,V)$ is the internal energy (obtained from `static' calculation) and vibrational contribution Gibb's free-energy. $G^{mag}_{HS/LS}(T,V)$ is the magnetic contribution to free-energy, which is given by
\begin{eqnarray}
 G^{mag}_{HS}(T,V) &=& -xk_BTln\left[m_{HS}(S_{HS}+1)\times m_{HS}(S_{HS}+1)\right], and \\ \nonumber
 G^{mag}_{HS}(T,V) &=& -xk_BTln\left[m_{HS}(S_{HS}+1)\times m_{LS}(S_{LS}+1)\right],
\end{eqnarray}
where  S$_{HS/LS}$ is spin (S$_{HS}$ = 5/2, S$_{LS}$ = 1/2) and m$_{HS/LS}$ is orbital degeneracy (m$_{HS}$ = 1, m$_{LS}$ = 3) of HS/LS state of [Fe$^{3+}$]$_{Si}$.

Minimizing $G(P,T,n,\sigma)$ (Eq.~\ref{G_MIX}) with respect to LS fraction $n(P,T,\sigma)$, we obtain
\begin{eqnarray}
  n(P,T,\sigma)= \frac{1}{1+\frac{m_{HS}(2S_{HS}+1)}{m_{LS}(2S_{LS}+1)}\exp\left\{\frac{\Delta G_{LS\rightarrow HS}}{xk_BT}\right\}},
\end{eqnarray}
 where $\Delta G_{LS\rightarrow HS} = G^{stat+vib}_{LS} - G^{stat+vib}_{HS}$. 

The elastic compliances $S^{ij}$ of the system in MS state are calculated as
\begin{eqnarray}
 S^{ij} = \left. -\frac{1}{V}\frac{\partial^2 G}{\partial\sigma_{i}\partial\sigma_{j}} \right|_{P,T}.
\end{eqnarray}
The derivative of $G$ with respect to $\sigma$ is computed from Eq.~(\ref{G_MIX})
\begin{eqnarray}
 \frac{\partial G}{\partial \sigma_i} = n\frac{\partial G_{LS}}{\partial \sigma_i} + (1-n)\frac{\partial G_{HS}}{\partial \sigma_i} + (G_{LS} - G_{HS})\frac{\partial n}{\partial \sigma_i} + \frac{\partial G^{mix}}{\partial \sigma_i}.
\end{eqnarray}
However, at equilibrium
\begin{eqnarray}\label{dG_dn}
  \frac{\partial G}{\partial n} = \left[(G_{LS} - G_{HS})\frac{\partial n}{\partial \sigma_i} + \frac{\partial G^{mix}}{\partial \sigma_i}\right]\frac{\partial \sigma_i}{\partial n} = 0. 
\end{eqnarray}
In Eq.~(\ref{dG_dn}), the derivative $\frac{\partial \sigma_i}{\partial n}\ne0$ ensures that
\begin{eqnarray}
 (G_{LS} - G_{HS})\frac{\partial n}{\partial \sigma_i} + \frac{\partial G^{mix}}{\partial \sigma_i} = 0.
\end{eqnarray}
Therefore, the elastic compliances for the system in MS state are 
\begin{eqnarray}\label{Sij_MIX}
 S^{ij}V = nS^{ij}_{LS}V_{LS} + (1-n)S^{ij}_{HS}V_{HS} + \left(\frac{\partial G_{HS}}{\partial \sigma_i} - \frac{\partial G_{LS}}{\partial \sigma_i}\right)\frac{\partial n}{\partial \sigma_j},
\end{eqnarray}
where volume $V$ in MS state is calculated using $V = nV_{LS} + (1-n)V_{HS}$.  The last term in Eq.~(\ref{Sij_MIX}) is responsible for the elastic anomalies due to iron spin crossover. 
Using the relation
\begin{eqnarray}
 \frac{\partial}{\partial \sigma_i} \equiv \frac{\partial}{\partial P} \frac{\partial P}{\partial \sigma_i} = \frac{1}{3}\frac{\partial}{\partial P},
\end{eqnarray}
for $i = 1-3$ and  $\frac{\partial n}{\partial \sigma_4} = \frac{\partial n}{\partial \sigma_5} = \frac{\partial n}{\partial \sigma_6} = 0$ for orthorhombic symmetric system, the elastic compliances are:
\begin{eqnarray}
 S^{ij}V = nS^{ij}_{LS}V_{LS} + (1-n)S^{ij}_{HS}V_{HS} - \frac{1}{9}(V_{LS} - V_{HS})\frac{\partial n}{\partial P} 
\end{eqnarray}
for $i,j = 1-3$ and
\begin{eqnarray}
 S^{ii}V = nS^{ii}_{LS}V_{LS} + (1-n)S^{ii}_{HS}V_{HS}
\end{eqnarray}
for $i = 4-6$. The elastic constants, $C_{ij}$, are  obtained using the relation: $\textbf{C}_{ij} = \left[\textbf{S} ^{-1} \right]^{ij}$.

%
 
\textbf{Acknowledgments}
This work was supported primarily by grants NSF/EAR 1319368, 1348066, and NSF/CAREER 1151738. 
Computations were performed at the Minnesota Supercomputing Institute (MSI) and at the Blue Waters System at NCSA. Results produced in this study are available in the Supporting Information.

\end{document}